\documentclass[10pt,conference,final]{IEEEtran}
\IEEEoverridecommandlockouts
\usepackage{amsmath,amssymb,amsfonts}
\usepackage{algorithmic}
\usepackage{graphicx}
\usepackage{textcomp}
\usepackage{xcolor}
\def\BibTeX{{\rm B\kern-.05em{\sc i\kern-.025em b}\kern-.08em
    T\kern-.1667em\lower.7ex\hbox{E}\kern-.125emX}}

\usepackage[hidelinks, bookmarks=false]{hyperref}

\usepackage{algorithm}
\usepackage{dsfont}
\usepackage{lineno}
\usepackage[utf8]{inputenc}
\usepackage[T1]{fontenc}
\usepackage[flushleft]{threeparttable}
\usepackage{xspace}
\usepackage{amssymb}
\usepackage{amsthm}
\usepackage{lscape}
\usepackage{url}
\usepackage{multirow}
\usepackage{array}
\usepackage{paralist}
\usepackage{makecell}
\usepackage[framemethod=tikz]{mdframed}
\newtheorem{definition}{Definition}

\usepackage{graphicx}
\usepackage{tabu}
\usepackage{booktabs}

\usepackage{color, colortbl}
\definecolor{Gray}{rgb}{220,220,220}

\usepackage{listings}
\mdfdefinestyle{mpdframe}{
    frametitlebackgroundcolor   =black!15,
    frametitlerule              =true,
    roundcorner                 =5pt,
    middlelinewidth             =0.8pt,
    innermargin                 =0.2cm,
    outermargin                 =0.2cm,
    innerleftmargin             =0.2cm,
    innerrightmargin            =0.2cm,
    innertopmargin              =0.2cm,
    innerbottommargin           =0.2cm
}

\usepackage{tikz}
\usetikzlibrary{arrows}
\usetikzlibrary{shapes}
\usetikzlibrary{patterns}
\usetikzlibrary{positioning}

\usepackage[colorinlistoftodos,prependcaption]{todonotes}
\newboolean{showcomments}
\setboolean{showcomments}{true}
\ifthenelse{\boolean{showcomments}}
 { \newcommand{\mynote}[2]{
      \fbox{\bfseries\sffamily\scriptsize#1}
        {\small$\blacktriangleright$\textsf{\textcolor{red}{{\em #2}\bf }}$\blacktriangleleft$}}}
        { \newcommand{\mynote}[2]{}}
\usepackage{tabularx}
\newcolumntype{C}{>{\centering\arraybackslash}X}
\newcolumntype{R}{>{\raggedleft\arraybackslash}X}

\usepackage{booktabs}

\definecolor{mymauve}{rgb}{0.58,0,0.82}
\definecolor{mygrey}{rgb}{0.43, 0.5, 0.5}

\usepackage{pbox}

\makeatletter
\newcommand*{\centerfloat}{%
  \parindent \z@
  \leftskip \z@ \@plus 1fil \@minus \textwidth
  \rightskip\leftskip
  \parfillskip \z@skip}
\makeatother 

\input{utils/utils.tex}

\usepackage{cite}

\usepackage{tikz}
\usetikzlibrary{arrows.meta}
\usetikzlibrary{arrows}
\usetikzlibrary{shapes}
\usetikzlibrary{patterns} 
\usetikzlibrary{positioning}
\usetikzlibrary{calc}

\newtheorem{metric}{Metric}

\begin{document}


\title{The Emergence of Software Diversity in \\ Maven Central
}

\author{
\IEEEauthorblockN{C\'esar Soto-Valero\IEEEauthorrefmark{2}, Amine Benelallam\IEEEauthorrefmark{1}, Nicolas Harrand\IEEEauthorrefmark{2}, Olivier Barais\IEEEauthorrefmark{1}, and Benoit Baudry\IEEEauthorrefmark{2}}
\IEEEauthorblockA{
\IEEEauthorrefmark{2}\textit{KTH Royal Institute of Technology, Stockholm, Sweden}\\ Email: \{cesarsv, harrand, baudry\}@kth.se\\
\IEEEauthorrefmark{1}\textit{Univ Rennes, Inria, CNRS, IRISA, Rennes, France}\\ Email: amine.benelallam@inria.fr, barais@irisa.fr
}
}
\maketitle


\begin{abstract}
Maven artifacts are immutable:  an artifact that  is uploaded on  Maven Central cannot be removed nor modified. The only way for developers to upgrade their library is to release a new version. Consequently, Maven Central accumulates all the versions of all the libraries that are published there, and applications that declare a dependency towards a library can pick any version. In this work, we hypothesize that the immutability of Maven artifacts and the ability to choose any version  naturally support the emergence of software diversity within Maven Central. We analyze 1,487,956 artifacts that represent all the versions of 73,653 libraries. We observe that more than 30\% of libraries have multiple versions that are actively used by latest artifacts. In the case of popular libraries, more than 50\% of their versions are used. We also observe that more than 17\% of libraries have several versions that are significantly more used than the other versions. Our results indicate that the immutability of artifacts in Maven Central does support a sustained level of diversity among versions of libraries in the repository.
\end{abstract}


\section{Introduction}\label{sec:introduction}

 

\mc is the most popular repository to distribute and reuse JVM-based artifacts (i.e., reusable software packages implemented in Java, Clojure, Scala or other languages that can compile to Java bytecode). By September 6, 2018, \mc contains over $2.8M$ artifacts and serves over $100M$ downloads every week~\cite{Benelallam2019}. The Maven dependency management system, which is able to resolve transitive dependencies automatically, has been key to this success: it relieves developers from the complexity of manual management of their dependencies. Uploading artifacts into \mc is the most effective way for open source projects to remain permanently accessible to their users. In this way, every build tool able to download Java libraries can fetch from a world of libraries and dependencies in a single and authoritative place.

In this work, we analyze software artifacts from the perspective of one essential characteristic enforced by \mc: immutability\footnote{Sonatype community support: \url{https://issues.sonatype.org/browse/OSSRH-39131}}. All artifacts (code packages, documentation, dependency declarations, etc.) that are uploaded on \mc are immutable: they cannot be rewritten nor deleted. This is a critical design choice that has a significant influence on the way the \mc repository is utilized. We hypothesize that this design decision is a great opportunity to prevent dependency monoculture~\cite{Stamp2004:Risks_of_Monoculture} and increase the diversity~\cite{Baudry2015:The_Multiple_Facets_of_Software_Diversity} among software dependencies. 

Previous works have analyzed Maven artifacts from the perspective of the risks induced by immutability. First, the redundancy in multiple versions can introduce conflicts among dependencies, e.g., trying to load the same class several times. This risk has been extensively analyzed by Wang and colleagues \cite{wang2018dependency}. 
Second, the projects that depend on a library need to explicitly update their dependency descriptions in order to benefit from the update. This represents a risk since these projects can eventually rely on outdated dependencies \cite{Kula2018:Do_Developers_Update_Their_Library_Dependencies} that
can contain security issues \cite{Pashchenko2018:Vulnerable_Open_Source_Dependencies} or API breaking changes~\cite{Jezek2015:How_Java_APIs_Break}.\looseness=-1

We take a fresh look at the presence of multiple versions of the same library in \mc, and consider it as an opportunity. 
We analyze how the ability to choose any library version for software reuse supports the emergence of software diversity in the repository and how this diversity of versions fuels the success of popular libraries. We consider this emergent diversity of reused versions as an opportunity since it participates in mitigating the risks of software monoculture \cite{lala2009:monoculture_risks}. Overcoming this type of monoculture is essential to build resilient and robust software systems~\cite{Gashi2007, Carzaniga2015,Baudry2015:The_Multiple_Facets_of_Software_Diversity}.

To conduct this empirical study, we rely on an existing dataset, the Maven Dependency Graph~\cite{Benelallam2019}, which captures a snapshot of \mc as of September 6, 2018. This dataset comes in the form of a temporal graph with metadata of $2.4M$ artifacts belonging to $223K$ libraries, with more than $9M$ direct dependency relationships between them. In order to enable reasoning not only at the artifact level but also at the library level, we extend this dataset with another abstraction layer capturing dependencies at the library level.




We measure activity, popularity and timeliness of a subset of $73,653$ libraries with multiple versions, which represents $61.81\%$ of the total number of artifacts in \mc. We empirically investigate whether the diversity of library versions is a valuable design choice. Our contributions are as follows:

\begin{itemize}
    \item a quantitative analysis of the diversity of usage and popularity of library versions;
    \item evidence of the presence of large quantities of artifacts that participate in the emergence of diversity;\looseness=-1
    \item open science with replication code and scripts available online.
\end{itemize}

\section{Background and Definitions}\label{sec:dataset}

In this section, we describe the dataset of Maven artifacts that constitutes the raw material for our work, as well as its extended library-level abstraction. 

\subsection{The Maven Dependency Graph}\label{subsec:mdg}

To conduct this empirical study, we rely on the Maven Dependency Graph (\mdg), a dataset that captures all of the artifacts deployed on the Maven Central repository as of September
6, 2018~\cite{Benelallam2019}. 
The \mdg includes $2,407,335$ artifacts. Each artifact is uniquely identified with a triplet (\textit{`groupId:artifactId:version'}). The \textit{groupId} identifier is a way of grouping different Maven artifacts, for instance by library vendor. The \textit{artifactId} identifier refers to the library name. Finally, the \textit{version} identifies each library release uniquely. 
For example, the triplet \textit{`org.neo4j:neo4j-io:3.4.7'} identifies the version 3.4.7 of an input/output abstraction layer for the Neo4j graph database.
The \mdg also includes  $9,715,669$ dependency relationships as declared in the Project Object Model (\texttt{pom.xml}) file of each artifact.

In this work, we focus on \textit{libraries}, i.e., the sets of artifacts that share the same tuple `\textit{groupId:artifactId}' but have different versions.  The \mdg includes $223,478$ of such libraries, but the concept of library is not rigorously captured in the graph. Consequently, we extend the artifact nodes of the \mdg with labels referring to their corresponding library. We call  \lib the set of all libraries in  \mc. We introduce an ordering function denoted  $\orderV$ that leverages the standard version numbering policy described by the Apache Software Foundation\footnote{\url{https://cwiki.apache.org/confluence/display/MAVEN/Version+number+policy}} in order to compare the different versions of artifacts belonging to the same library. For instance, 1.2.0 $\orderV$ 2.0.0. We also define a temporal ordering function denoted by $\orderT$ to compare the release dates of different artifacts. For example, `12-09-2011' $\orderT$ `30-03-2015'.
In the remainder of the paper, we refer to artifacts as \textit{library versions} or simply \textit{versions}. We define the \mdg as follows:

\begin{definition} \label{def:mdg} 
\textbf{Maven Dependency Graph.} The \mdg is a vertex-labelled graph, where vertices represent Maven library versions, and edges describe dependency relationships or precedence relationships. We use a labeling function over vertices to group versions by library. We define the \mdg as  $\mathcal{G} = (\mathcal{V}, \mathcal{D}, \mathcal{N}, \mathcal{L},\mathcal{R})$, where,

\begin{figure}[htb]
	\centering
	\includegraphics[origin=c,width=0.95\columnwidth]{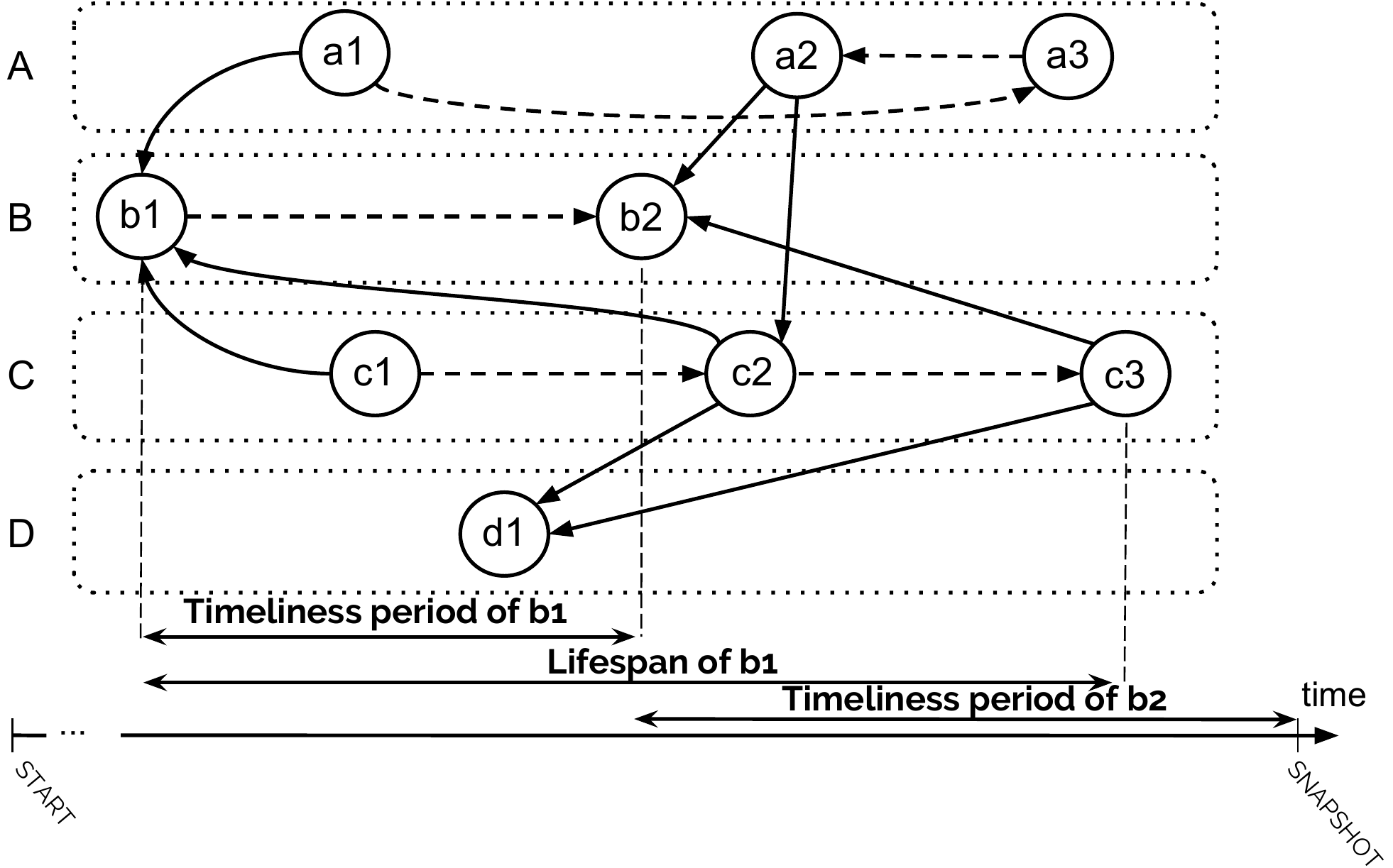}
	\caption{Example of relationships between library versions in the Maven Dependency Graph.}
	\label{fig:graph_example}
\end{figure}

\begin{itemize}
    \item the set of vertices $\mathcal{V}$ represents the library versions present in \mc
    \item the set of directed edges  $\mathcal{D}$  represents dependency relationships between library versions
    \item the set of directed edges $\mathcal{N}$ represents versions precedence relationships, where the version of the source node is strictly lower than the version of the target node \wrt $\orderV$\looseness=-1
    \item the surjective labelling function $\mathcal{L}$ returns the corresponding library of a given library version $v \in \mathcal{V}$, defined as  $\mathcal{L}: \mathcal{V}\rightarrow\text{\lib}$
    \item the temporal function $\mathcal{R}$ refers to the date at which a library version $v \in \mathcal{V}$ was deployed, defined as $\mathcal{R} : \mathcal{V} \rightarrow T$, where $T$ is a $\orderT$-ordered discrete time domain
\end{itemize}
\end{definition}

In the \mdg, $T$ is bounded to [`15-05-2002', `06-09-2018'], where the lower bound refers to the date when the first library was deployed on \mc. In the rest of the paper, we refer to the lower and upper bounds respectively as \start and \snapshot, and we use days as the time granularity. 

Figure~\ref{fig:graph_example} illustrates the different nodes and relationships within a simplified graph $\mathcal{G}$ composed of four libraries (A,B,C,D) and nine library versions  $\{a_1, a_2, a_3, b_1, b_2, c_1, c_2, c_3, d_1\}$.  The regular edges represent dependency relationships. For example, the first version of A ($a_1$) depends on the first version of B ($b_1$), and the second version of A ($a_2$) depends on the second version of B ($b_2$) and C ($c_2$). The dashed edges represent precedence relationships, and all vertices that are related through such edges constitute the different versions of a library. In Figure~\ref{fig:graph_example}, we place nodes in a temporal order, from left to right, corresponding to the deployment date, thus the node $b_1$ is the firstly deployed, while the node $c_3$ is the most recently deployed.

The temporal order of releases does not imply a similar versioning order for a given library. In some cases, library instances with lower version number may be released after library versions with a greater version number, e.g., in case of a library version downgrade or maintenance of several major library versions. In Figure~\ref{fig:graph_example}, we can see that $a_2 \orderT a_3$ and $a_3 \orderV a_2$. 
Note, this is a common practice adopted by very popular libraries such as Apache CXF\footnote{\url{https://cxf.apache.org}}, and Mule\footnote{\url{https://www.mulesoft.com}}~\cite{Suwa2017:An_Analysis_of_Library_Rollbacks}.

\begin{definition} \label{def:shortcuts} 
\textbf{Additional notations.}
For further references in the \mdg, we introduce the following  notations:
\begin{itemize}
    \item $next(v)$: the next release of a given library version $v$ \wrt the ordering function $\orderV$
    \item $next_{all}(v)$: transitive closure on the next releases of a library version $v$
    \item \latest: the library version $v$ such that $ \nexists$ $next(v)$
    \item \latests: the set of all latest library versions in a dependency graph $\mathcal{G}$
    \item  $deps(v)$: $\mathcal{V} \rightarrow \mathcal{V}^n,$ with $n \in \mathds{N}$: the set of direct dependencies of a given library version $v \in \mathcal{V}$ 
    \item $deps_{tree} (v)$: the whole dependency tree of $v$
    \item  $users(v)$: $\mathcal{V} \rightarrow \mathcal{V}^n,$ with $n   \in \mathds{N}$: the set of library versions  declaring a dependency towards $v$
    \item $users_{all}(v)$: all the transitive users of $v$
\end{itemize}
\end{definition}

For example, in Figure~\ref{fig:graph_example}, $deps(a_2)=\{b_2, c_2\}$, $deps_{tree}(a_2) =\{b_2, c_2, d_1\}$, $users(d_1)=\{c_2,c_3\}$ and $users_{all}(d_1)=\{c_2,c_3,a_2\}$.\looseness=-1

\subsection{The Maven Library's Dependency Graph}\label{subsec:the maven library dependency graph}

\begin{figure}[htb]
	\centering
	\includegraphics[origin=c,width=0.72\columnwidth]{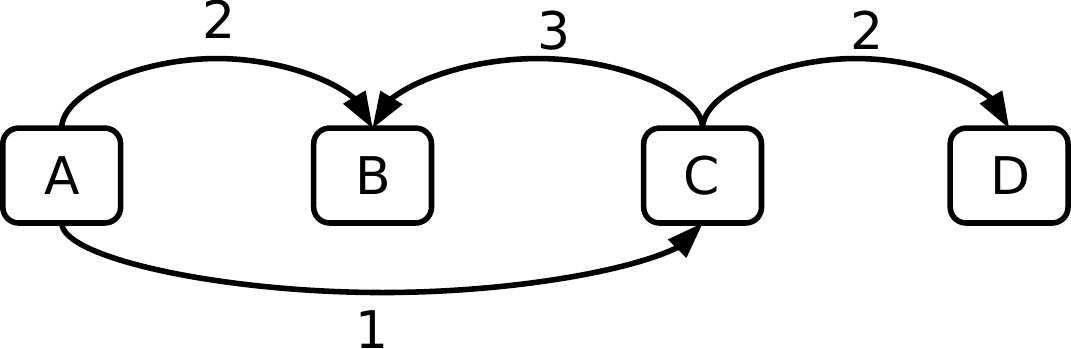}
	\caption{The elevated Maven Library's Dependency Graph from Figure~\ref{fig:graph_example}.}
	\label{fig:libs_graph}
\end{figure}

In order to be able to reason about not only versions but also libraries, we elevate the abstraction of the \mdg to the library level.
Figure~\ref{fig:libs_graph} shows the elevated graph corresponding to the dependency graph $\mathcal{G}$ in Figure~\ref{fig:graph_example}.
We construct a weighted graph, $\mathcal{G}_\mathcal{L}$, where nodes correspond to libraries (\lib) in $\mathcal{G}$. We create an outgoing edge between two libraries $l_1$ and $l_2$ if there is at least a version of $l_1$ that uses a library version of $l_2$. We denote by $D(l)$ the set of direct library dependencies of a given library $l$. For example, $D(A)$ = \{$B$, $C$\}. Finally, the weight of the outgoing edges from $l_1$ to $l_2$ corresponds to the number of versions of $l_1$ that use a version of $l_2$. We define the Maven Library’s Dependency Graph (\mdgl) as follows:\looseness=-1

\begin{definition}\label{def:Maven-libraries-dependency-graph}
\textbf{Maven Library’s Dependency Graph.} The \mdgl is a edge-weighted graph, where vertices represent Maven libraries, and edges' weight describes the number of dependency relationships between their versions. We define the \mdgl as $\mathcal{G}_\mathcal{L}$ = (\lib, $\mathcal{E}$, $\mathcal{W}$), where,
\begin{itemize}
    \item the set of vertices \lib represents the libraries present in Maven Central
    \item the set of edges $\mathcal{E}$ represents the dependency relationships between libraries
    \item the weighing function $\mathcal{W}$ represents the weight of a given edge, defined as  $\mathcal{W} : \mathcal{E}\rightarrow \mathds{N}$
\end{itemize}
For further references in the \mdgl, we introduce the following notations:
\begin{itemize}    
    \item the set of direct library dependencies $D$ of a given library, defined as $D :$ \lib $\rightarrow \text{\texttt{LIBS}}^{n}$
    \item the weighing function $\overleftarrow{W}$ returns the sum of the weights of incoming edges, defined as   $\overleftarrow{W} :$ \lib $\rightarrow \mathds{N}$
    \item the weighing function $\overrightarrow{W}$ returns the sum of the weights of outgoing edges, defined as  $\overrightarrow{W} :$\lib $\rightarrow \mathds{N}$
\end{itemize}
\end{definition}

\section{Study design}\label{sec:study_design}

This work is articulated around five research questions. In this section, we introduce these questions as well as the metrics that we collect to answer them. We also describe the representative subset of artifacts that we study throughout the paper.\looseness=-1




\subsection{Research Questions}\label{sec:research_questions}

\newcommand{\RQone}{\textbf{RQ1}: To what extent are the different library versions actively used?}
\subsection*{\RQone}

Because Maven artifacts are immutable, all the versions of a given library that have been released in \mc are always present in the repository. Meanwhile, previous studies have shown that users of a given version do not systematically update their dependency when a new version is released~\cite{Kula2018:Do_Developers_Update_Their_Library_Dependencies, Kula2015:Trusting_a_Library, Bavota2015:How_the_Apache_Community_Upgrades_Dependencies}. Consequently, we hypothesize that, at some point in time, multiple versions of a library are actively used. In this research question, we investigate how many versions are currently used, how many have been used but are not anymore and how many versions have never been used.

\newcommand{\RQtwo}{\textbf{RQ2}: How are the actively used versions distributed along the history of a library?}
\subsection*{\RQtwo}

The full history of versions of a library released on \mc is always available. Consequently, users can decide to depend on any of the versions. In this research question, we analyze where, in the history of versions, are located the versions that are actively used. 

\newcommand{\RQthree}{\textbf{RQ3}: Among the actively used versions of a library, is there one or several versions that are significantly more popular than the others?}
\subsection*{\RQthree} 

Library users are free to decide which version to depend on and for how much time. In the long term, these users' decisions determine what are the most popular libraries and versions in the entire software ecosystem~\cite{Kula2018:Do_Developers_Update_Their_Library_Dependencies, Mileva2009:Mining_Trends_of_Library_Usages}. This research question investigates to what extent these decisions lead to the emergence of one or more versions that receive a greater number of usages compared to the other versions.


\newcommand{\RQfour}{\textbf{RQ4}: Does the number of actively used versions relate to the popularity of a library?}
\subsection*{\RQfour}

We observe that for most libraries, more than one version is actively used at a given point in time. The library developers have no control over this since they cannot remove versions from \mc, nor force their users to update their dependencies. Meanwhile, it might be good for a library to maintain several versions that fit different usages. In this question, we investigate how the existence of multiple active versions relates to the overall popularity of a library. 

\newcommand{\RQfive}{\textbf{RQ5}: How timely are the different library versions in Maven Central?}
\subsection*{\RQfive}
With each new release, project maintainers make an effort to improve the quality of their libraries (\eg, by fixing bugs, adding new functionalities or increasing performance). These changes are expected to be directly reflected in the number of users that update their dependencies to the new available release, and also in the number of new usages of the library~\cite{Kula2018:An_Exploratory_Study_on_Library_Aging_by_Monitoring_Client_Usage}. This research question aims to get insights into how timely is the release of new versions. In particular,  we investigate how much attraction gets a library version while it was the latest, compared to the older versions during the same period of time.

\subsection{Metrics}

To characterize the \textit{activity status} of libraries and versions in terms of their usages by other latest library versions, we introduce the notions of \textit{active}, \textit{passive}, and \textit{dormant} libraries and versions. Moreover, we introduce the \textit{lifespan} of library versions to get insights on the duration of their activity period. These notions and measures are intended to answer RQ1 and RQ2.\looseness=-1

\begin{metric}\label{met:passive}
\textbf{Activity status.} 
A \textbf{passive library version} $v$ is a version that has been used in the past, but is no longer used, even transitively, by any latest library version ($v \in$  \latests). 
Formally, this metric is described as a boolean function $\isPassive : \mathcal{V} \rightarrow \{true, false\}$, where,
\begin{equation*}
\isPassive (v) = \begin{cases}
false & v \in \displaystyle \bigcup_{i \in \text{\latests}} \{ deps_{tree}(i)\}\\
true, &\text{otherwise}
\end{cases}
\end{equation*}

An \textbf{active library version} $v$ is a version where $isPsv(v) = false$. 
A \textbf{dormant library version} is an extreme case of a passive library version that occurs when the version has never been used by existing libraries  (\ie, $users(v) = \emptyset$) in \mc.\looseness=-1


At the library level, an \textbf{active library} is a library that has at least one active version, whereas a \textbf{passive library} is a library that has all its versions passive. A \textbf{dormant library} is an extreme case of passive library that occurs when all its versions are dormant.
\end{metric}

\begin{metric}\label{def:lifespan}
\textbf{Lifespan.} The lifespan of a library version $v$ is the time range during which it was/is being used. We define this period as the time range between the release date of $v$ and the timestamp at which it becomes passive. In case $v$ is active, this period starts at the release date of the artifact until the day the \snapshot was captured. Dormant library versions do not have a lifespan at all. We denote this metric by $ls (v) = $[$startLs_v,endLs_v$].  Then, the interval's upper bound can be formally described as follows: \looseness=-1

\begin{equation*}
endAct_v = 
\begin{cases}
\text{\snapshot}, & \neg \isPassive(v)  \\
 last, & \isPassive(v)
\end{cases}
\end{equation*}
where, $last =  \max \displaystyle \bigcup_{i \in users_{all}(v)} \{\mathcal{R}(next(i)) \}$.
\end{metric}


To study the \textit{popularity} of library versions in \mc, and hence answer RQ3 and RQ4, we introduce a metric of popularity which measures the transitive influence and connectivity of a library version in the \mdg.  We rely on the standard PageRank algorithm~\cite{Page1999:The_PageRank_Citation_Ranking}, which accounts for the number of transitive usages. Intuitively, library versions with a higher PageRank are more likely to have a larger number of transitive usages. 
On the other hand, to measure the popularity of libraries, we use the Weighted PageRank algorithm~\cite{Xing2004:Weighted_PageRank_Algorithm} on the \mdgl.
\begin{metric}\label{met:popularity-ver}
\textbf{Popularity.}
The \textbf{popularity of a library version} $v \in \mathcal{V}$ is as follows:
\begin{equation*}
pop_\mathcal{V}(v) = (1-d) + d \displaystyle \sum_{i \in users(v)} pop_\mathcal{V}(i),
\end{equation*}
where $d$ is a damping factor to reflect user behavior, which is usually set to 0.85~\cite{Boldi2005:PageRank_As_a_Function_of_the_Damping_Factor}.

The \textbf{popularity of a library} $l \in \text{\lib}$ is as follows: 
\begin{equation*}
pop_\mathcal{L}(l) = (1-d) + d \displaystyle \sum_{i \in U(l)}pop_{v}(i) \overleftarrow{c}_{(l,i)} \overrightarrow{c}_{(l,i)}, 
\end{equation*}

where $\overleftarrow{c}$ and $\overrightarrow{c}$ are respectively:
\begin{equation*}
\overleftarrow{c}_{(l,i)}=\frac{\overleftarrow{W}(i)}{\displaystyle \sum_{p \in D(l)}\overleftarrow{W}(p)}, 
\hspace{0.3cm}
\overrightarrow{c}_{(l,i)}=\frac{\overrightarrow{W}(i)}{\displaystyle \sum_{p \in D(l)}\overrightarrow{W}(p)}.\label{equ:ud}
\end{equation*}
\end{metric}


 Finally, to answer RQ5, we introduce  the notion of  \textit{timeliness} of library versions. This metric looks at the number of usages of every single version when it was latest and assesses if it was successful in attracting more users compared to its older versions. To this end, we compare the usages of a given version $v$ during its lifespan to the usages that the whole library has received during the period when $v$ was latest. We call this period the \textit{timeliness period}.

\begin{metric}\label{met:timeliness}
\textbf{Timeliness.} 
The \textbf{timeliness period}, $tp(v)$, of a  library version $v$, is the time range between the release date of $v$ and the most recently released version of its library ordered by $\orderT$, which is not necessarily $next(v)$. We denote this version as $mr$:

$tp(v) = [\mathcal{R}(v),\mathcal{R}(mr)]$, 

where, $mr =  \displaystyle \min_{i \in next_{all}(v)} \{\mathcal{R}(i) | \mathcal{R}(i) >_t \mathcal{R}(v)\}$.

The \textbf{timeliness of a library version} $v$ is a function, $tim(v): \mathcal{V}\rightarrow\mathds{Q}^+$, where, 
\begin{equation*}
    tim(v) = \frac{|users(v)|}{|\displaystyle \bigcup_{i \in \mathcal{V}} \{i | \mathcal{R}(i) \in tp(v) \wedge \mathcal{L}(v) \in {\displaystyle \bigcup_{j \in deps(i)} \{\mathcal{L}(j)\}\}}|}.
\end{equation*}

In case  the library corresponding to $v$ was not  used during the timeliness period of $v$ (the denominator is 0), then we consider $tim(v) = 0$. This also applies when $v$ is dormant. All first releases of libraries have $tim(v)=1$ since they have no earlier releases.

Based on the \textit{timeliness} metric, three situations can occur:
\begin{itemize}
    \item $v$ is \textbf{timely} if $tim(v) = 1$: $v$ was a success during its timeliness period and users relied on it
    \item $v$ is \textbf{over-timely} if $tim(v) > 1$: $v$ has attracted  users beyond its timeliness period
    \item $v$ is \textbf{under-timely}  if  $tim(v) < 1$: users relied on older versions during its timeliness period
\end{itemize}

\end{metric}


\subsection{Study Subjects} 

During our initial exploration of the \mdg, we distinguished three different categories of libraries in \mc: (i) libraries that have only one version (${\sim}30\%$), (ii) libraries with multiple versions all released on the same day (${\sim}15\%$),
and (iii) libraries with multiple versions released within different time intervals (${\sim}55\%$). Table~\ref{tab:versions} gives detailed numbers about these categories. In particular, after manual inspection we notice that a large number of libraries belonging to categories (i) and (ii) are shipped with their classpath. We suspect these projects to be using Maven only for deploying and storing their libraries in \mc, but not for dependency management or further maintenance tasks.

\begin{figure}[htb]
	\centering
	\includegraphics[origin=c,width=0.46\textwidth]{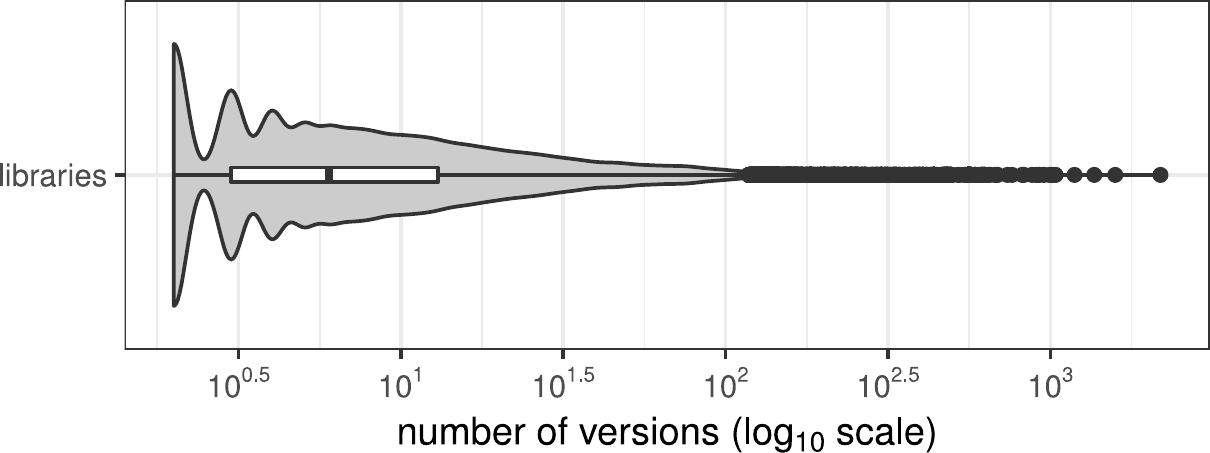}
	\caption{Distribution of the number of library versions in \mc.}
	\label{fig:version_distribution}
\end{figure}

In this work, we are interested in studying libraries that have multiple versions and utilize Maven regularly to manage and update their dependencies, \ie, libraries belonging to category (iii) in Table~\ref{tab:versions}. Figure~\ref{fig:version_distribution} shows the distribution of the number of versions for the libraries in this category. The minimum and maximum number of versions are respectively $2$ and more than $2,000$, precisely, $2,122$. Meanwhile, the 1st-Q and 3rd-Q are around $5$ and $200$ versions respectively. 

\begin{table}[!tbp]
 \begin{threeparttable}
\centering
\caption{Categories of libraries in \mc according to their releasing profiles}
\begin{tabular}{cccc} 
\toprule
\textbf{Category} & \textbf{Criteria} & \textbf{\#Libraries (\%)} & \textbf{\#Versions (\%)} \\ 
\hline
(i) & \#versions = $1$ & $65,557$ ($29.33\%$) & $65,557$ ($2.72\%$) \\
(ii) & One shot*  & $32,825$ ($14.69\%$) & $459,445$ ($19.08\%$) \\
(iii) & \#versions > $1$ & $125,096$ ($55.98\%$) &  $1,882,333$ ($78.2\%$)\\
\bottomrule

\label{tab:versions}
\end{tabular}
\begin{tablenotes}
      \footnotesize
      \item (*) Libraries with more than one version and that have been released in the same day.
    \end{tablenotes}
\end{threeparttable}

\end{table}

In order to conduct our empirical study on a representative dataset, we choose [1st-Q, 3rd-Q] as a range of number of versions. Therefore, this study focuses on all the libraries with between $5$ and $200$ versions. This accounts for $73,653$ libraries and $1,487,956$ versions, representing $32.96\%$ and $61.81\%$ of the total number of libraries and version in \mc at the \snapshot time.\looseness=-1


\section{Results}\label{sec:results}

In this section, we address our research questions and present the results obtained.

\subsection{\RQone}

To answer RQ1, we study the activity status of libraries and versions in \mc. Table~\ref{tab:act_pas} shows the numbers and percentages of active, passive and dormant libraries and versions. We observe a low percentage of active versions ($14.73\%$), whereas there is a predominant number of passive ones ($85.27\%$), of which more than a half are dormant ($45.16\%$). On the other hand, we can notice that the majority of libraries are active ($95.49\%$), i.e., have at least one of its versions active.  
Meanwhile, passive libraries represent nearly $5\%$ of the total number of libraries, of which (${\sim}4\%$) are dormant.

\begin{figure}[htb]
	\centering
	\includegraphics[origin=c,width=0.4\textwidth]{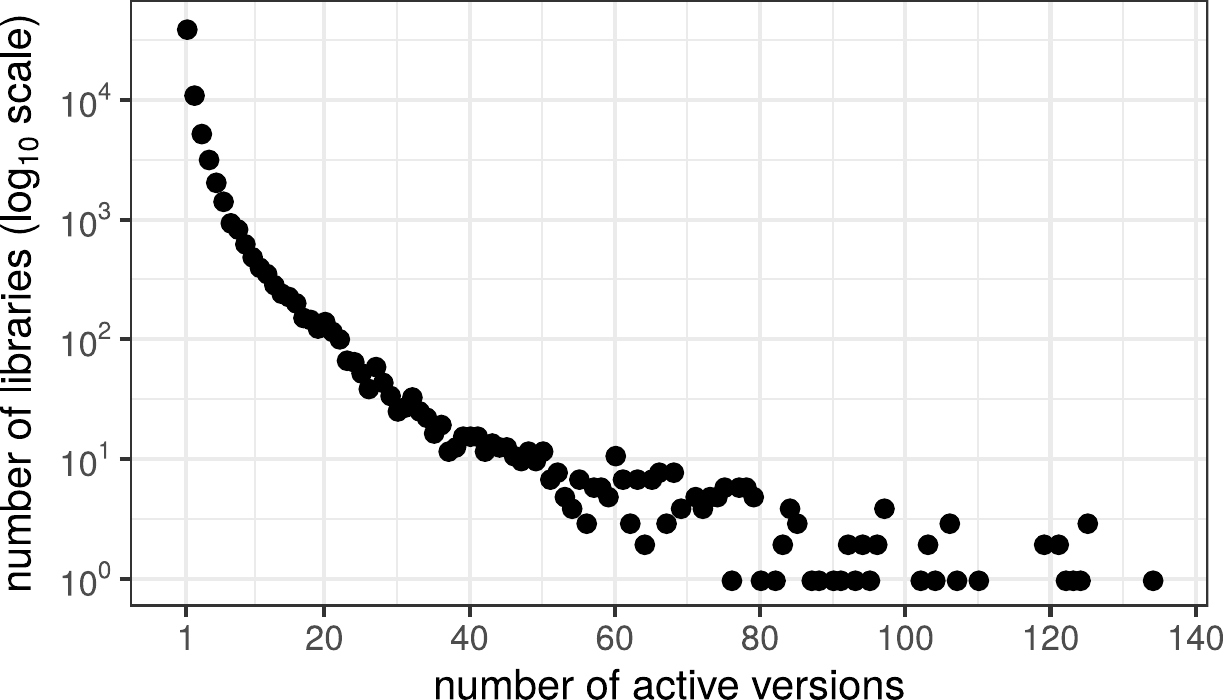}
	\caption{Distribution of the number of active versions across active libraries}
	\label{fig:numactlib_vs_numactver}
\end{figure}

\begin{table}[htb]
\centering
\caption{Activity status of libraries and versions in the study subjects}
\begin{tabular}{lcc} 
\toprule
\textbf{Status} & \textbf{\#Versions (\%)} & \textbf{\#Libraries (\%)} \\ 
\hline
Active  &  $219,184$ ($14.73\%$) & $70,337$ ($95.49\%$)\\
Passive non-dormant  &  $596,776$ ($40.11\%$) & $387$ ($0.53\%$)\\
Dormant  &  $671,996$ ($45.16\%$) & $2,929$ ($3.98\%$)\\ 
\hline
Total & $1,487,956$ ($100\%$) & $73,653$ ($100\%$)\\
\bottomrule
\label{tab:act_pas}
\end{tabular}
\end{table}

We are intrigued by the $2,929$  dormant libraries. The median number of versions in this family of libraries is $9$ with a maximum of $150$ versions. 
We noticed that most of them are in-house utility libraries, intended for custom logging or testing, \eg, `\emph{com.twitter:util-benchmark\_2.11.0}'. Other libraries are archetypes\footnote{\url{https://maven.apache.org/guides/introduction/introduction-to-archetypes}}, \eg, `\emph{io.fabric8.archetypes:karaf-cxf-rest-archetype}'.
These libraries are not intended to be used in production. Their custom nature makes them used rather internally, or by the library maintainers themselves. \looseness=-1


In Table~\ref{tab:act_pas}, we also observe that a low proportion of versions are active $219,184$ ($14.73\%$), yet they are distributed across a very high number of libraries, $70,337$ ($95.49\%$), making these libraries active. 
Figure~\ref{fig:numactlib_vs_numactver} summarizes the distribution of active versions in active libraries. We observe that more than a half of active libraries, $40,233$ in total, have only one active version. The remainder, $30,104$ libraries, have more than one active version. For some libraries, such as `\emph{org.hibernate:hibernate-core}', more than $100$ versions are currently active. However, the number of libraries with more than $100$ active versions represents less than $2\%$ of the total. More interestingly, we notice that $17\%$ of the libraries have active versions belonging to more than one different major releases (\eg, 2.X.X). For instance, the library `\emph{activemq:activemq}' has two active major versions: 3.X.X and 4.X.X, whereas `\emph{com.spotify:docker-client}' has seven active versions: from 2.X.X up to 8.X.X. 

\begin{figure}[htb]
	\centering
	\includegraphics[origin=c,width=0.38\textwidth]{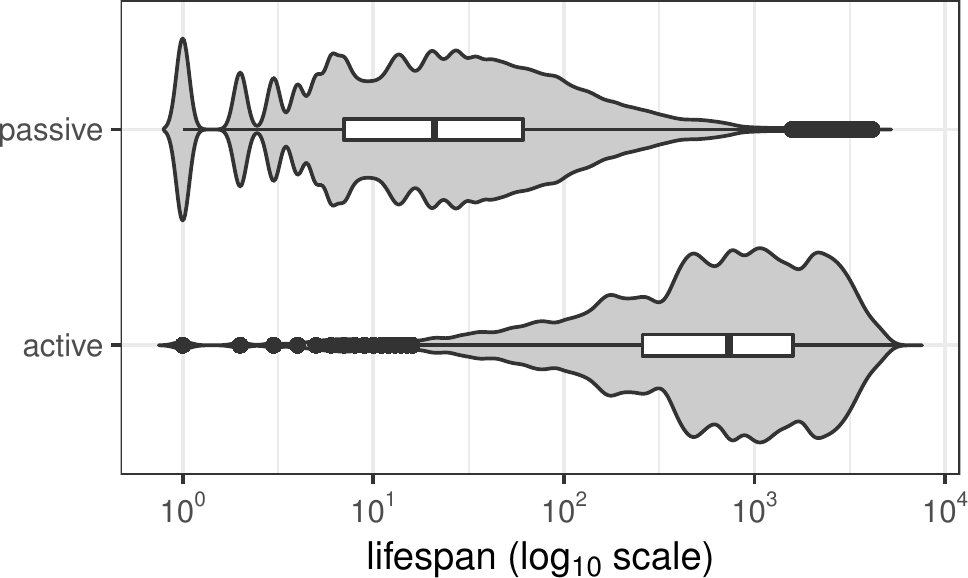}
	\caption{Distributions of the lifespan (in days) of passive and active  versions}
	\label{fig:dist_lifespan}
\end{figure}

Figure~\ref{fig:dist_lifespan} shows the lifespan distribution of active and passive versions. To avoid the bias introduced by the \snapshot time constraint, we consider only non-latest active versions of libraries ($v \notin \latests$). As we can see from the figure, the lifespan of passive versions is approximately distributed between $8$ and $80$ days (1st-Q and 3rd-Q), whereas, this range is larger for active versions: between $351$ and $1,626$ days. This conveys that versions that are active for more than $80$ days are likely to remain active for a  longer period. Subsequently, these libraries are likely to be popular and widely used. 
Finally, we notice that the median number of days a version spends after its creation before being used for the first time is $14$, with a mean of $57.61$. This suggests that versions that have been dormant for less than $57$ days are likely to become active; beyond this time period, they are likely to remain dormant. 

\begin{mdframed}[style=mpdframe]
\textbf{Findings from RQ1:}
More than $40\%$ of libraries in \mc have strictly more than one active version, while almost $4\%$ of the libraries have never been used. This  hints on an inclusive, immutable repository that can support the emergence of a diversity of library usages.
\looseness=-1   
\end{mdframed}

\subsection{\RQtwo}

According to Metric~\ref{met:passive}, active libraries have at least one active version. In this research question, we focus on understanding how these active versions are distributed across the different library releases.

\begin{figure}[htb]
	\centering
	\includegraphics[origin=c,width=0.4\textwidth]{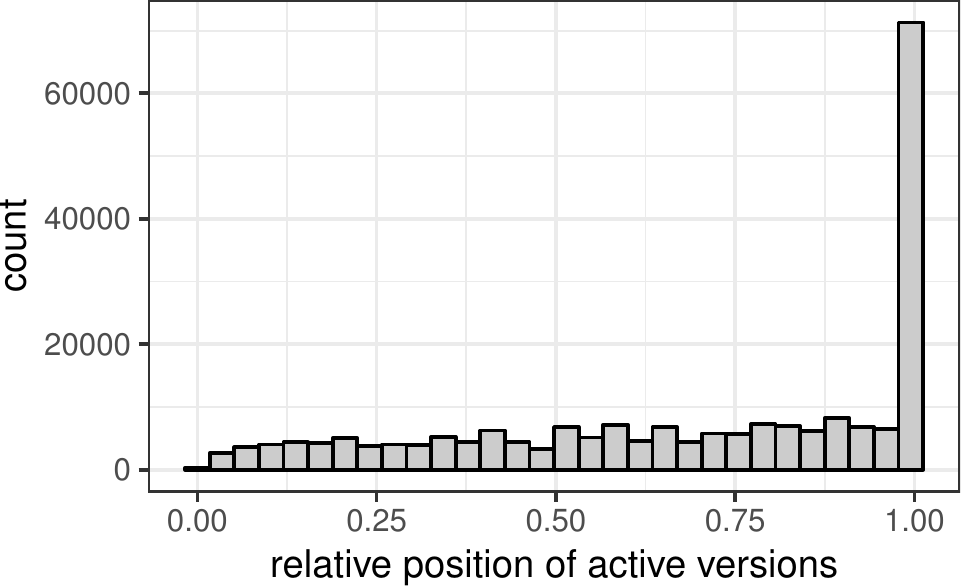}
	\caption{Positional distribution of active versions (\#bins = $30$).}
	\label{fig:hist_pos_active_versions}
\end{figure}

Figure~\ref{fig:hist_pos_active_versions} shows the positional distribution of all the active versions in the libraries. Since libraries can have different number of versions, we use a normalized relative index lying between $[0,1]$, where $0$ and $1$ represent the indexes of the first and last versions of the library, respectively. First of all, we observe that active versions are scattered across different positional indices. While $68.4\%$ of active library versions are almost evenly distributed across the non-latest releases, a significant number of active versions, precisely $69,146$ ($31.6\%$), are latest versions. This result is inline with the current policies of dependency management systems, which recommend upgrading to latest dependencies. 

Digging further, we investigate the transitional distribution of active and passive versions. To do this, we transform each library $l \in$ \lib into a vector, $\mathcal{S}_l$, capturing the passive/active status corresponding to all of its versions. Our objective is to analyze the occurrence of common transitional patterns between active and passive versions. 

Let $\mathcal{S}_l$ be a vector representing the activity status of library versions ordered by $\orderV$ (\ie ordered by version number). The status corresponding to a version $v$ is $\pas$ if $isPSV(v)$ is $TRUE$ and $\act$ otherwise.
For example, the library \emph{com.google.guava:guava-jdk5} has a total of five versions, i.e., $\mathcal{S}_{guava-jdk5}$ = [\act,\act,\pas,\pas,\act]. Considering that we are particularly interested in transitional patterns, the consecutive versions with the same status can be compressed to a single status, \eg, the previous example is represented as [\act,\pas,\act].\looseness=-1


\begin{table}[htb]
\centering
\caption{The top-$7$ most common transitional patterns}
\begin{tabular}{lcl} 
\toprule
\textbf{Pattern} & \textbf{Frequency} & \textbf{Example} \\ 
\hline
[\pas,\act]  &  $43,549$ & \emph{commons-codec:commons-codec} \\
\text{[\pas,\act,\pas,\act]}  &  $10,219$ & \emph{org.apache.commons:commons-lang3} \\
\text{[\pas,\act,\pas,\act,\pas,\act]}  &  $3,478$ & \emph{org.jboss.logging:jboss-logging} \\ 
\text{[\act,\pas, \act]}  &  $2,761$ & \emph{com.google.guava:guava-jdk5} \\ 
\text{[\pas,\act,\pas,\act,\pas,\act,\pas,\act]}  &  $1,592$ & \emph{org.joda:joda-convert} \\ 
\text{[\act,\pas,\act,\pas,\act]}  &  $1,343$ & \emph{com.google.inject:guice}\\ 
\text{[\pas,\act,\pas]}  &  $613$ & \emph{org.springframework:spring-webflow} \\ 
\bottomrule
\label{tab:patterns}
\end{tabular}
\end{table} 

We obtained a total of $94$ different transitional patterns. Table~\ref{tab:patterns} shows the frequency of appearance of the seven most common of them. As expected, the $92\%$ of the patterns are finishing by an \act. The most frequent pattern is [\pas,\act], i.e., old versions are  passive and the latest ones are active. Yet, the remaining patterns represent more than $40\%$ of the libraries. The rest of libraries follow a pattern where some old versions are also active. In extreme cases, the latest version of the library is passive (patterns finishing with a \pas \@\xspace). In such cases, we observe that most of their clients use an older version with the same major version number. We speculate that this behavior is due to the clients' belief that the version they use is rather stable. Similar findings have been reported by Kula~\etal~\cite{Kula2015:Trusting_a_Library}.
We also observe that $5.5\%$ of the libraries have their earliest version active. It is interesting to note that many of them are very popular libraries, \eg, `\textit{org.hamcrest:hamcrest-core}' and `\textit{org.apache.ant:ant}'.\looseness=-1




\begin{mdframed}[style=mpdframe]
\textbf{Findings from RQ2:} 
$31.6\%$ of active versions are latest and the remaining $68.4\%$ of active versions are evenly distributed across the  libraries' history.  When the clients do not use the latest version,  they often depend on earlier versions belonging to the same major release of the library.\looseness=-1
\end{mdframed}

\subsection{\RQthree}

In this research question, we investigate the diversity in the popularity of library versions. We assess the popularity of a library versions using Metric~\ref{met:popularity-ver}. In particular, we are interested in identifying significantly popular versions and  analyzing the positional distribution of these versions.   
For this aim, we use the Tukey's outlier detection method~\cite{Tukey1977:Exploratory_Data_Analysis} to identify versions with a popularity score that is far greater than the remaining versions of the library. 

We distinguish between three different classes of libraries: (i) libraries that do not have a significantly most popular version ($55,148$), (ii) libraries with one significantly popular version ($9,622$), and (iii) libraries with more than one significantly popular version ($8,883$). The first class (i) represents libraries with versions that have a similar number of usages. The classes (ii) and (iii) represent libraries with one or more versions that have attracted more users compared to the rest of their versions. A large number of the users of significantly popular versions are different versions of the same library. These are library providers that may have remained loyal to one version despite the release of newer versions. To our surprise, almost all the significantly popular versions are active, only $86$ out of $143,334$ are passive. For instance, `\emph{com.amazonaws:aws-java-sdk:1.11.409}' is significantly popular and passive.\looseness=-1

\begin{figure}[htb]
	\centering
	\includegraphics[origin=c,width=0.45\textwidth]{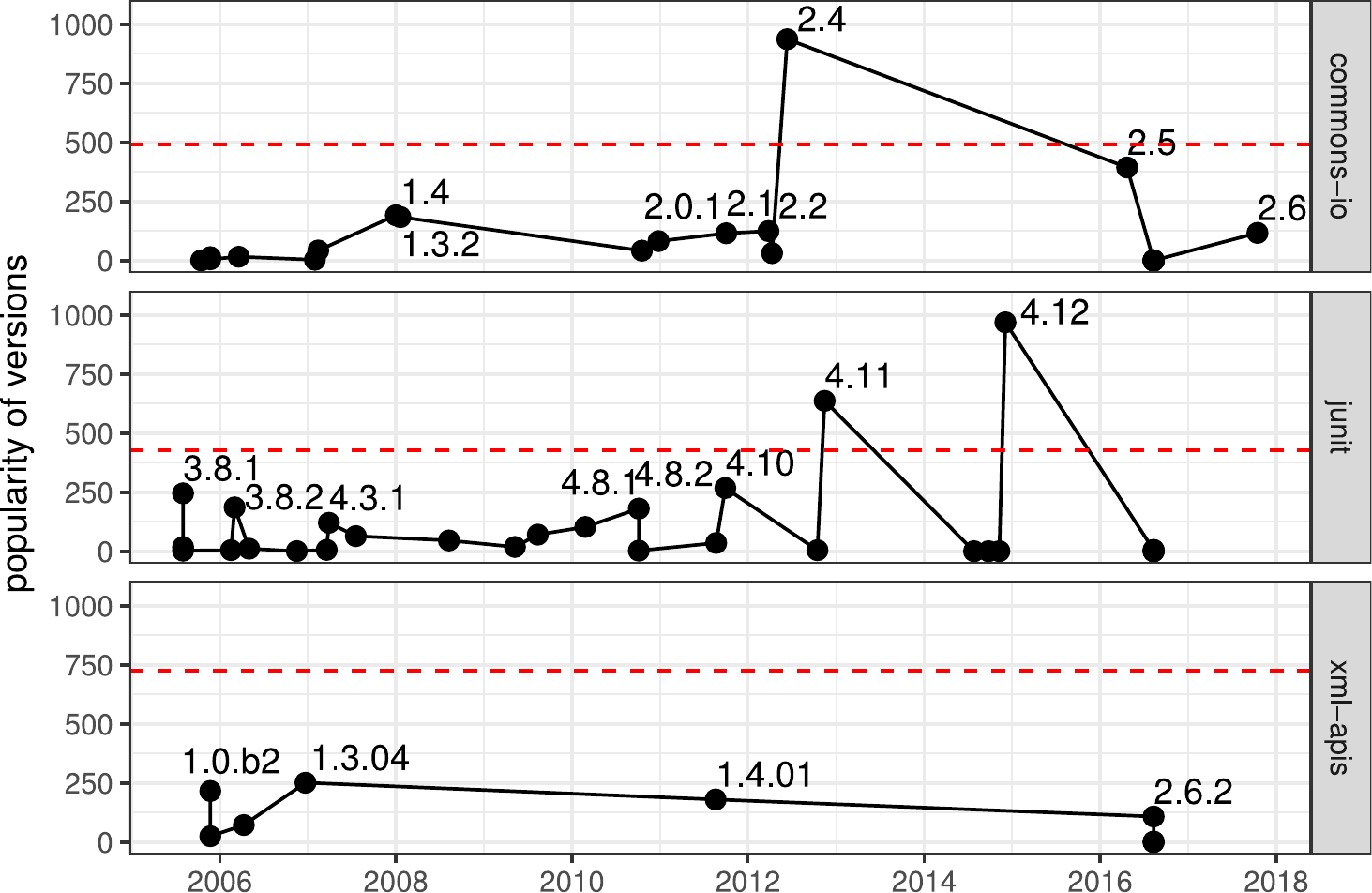}
	\caption{Evolution of the popularity of versions ($pop_\mathcal{V}(v)$ metric) corresponding to the libraries Apache Commons IO, JUnit and XML APIs.}
	\label{fig:examples_evolution}
\end{figure}

Figure~\ref{fig:examples_evolution} shows illustrative examples, Apache IO, JUnit, and XML APIs, each one corresponding to one of these three classes. The horizontal dashed line in each frame represents the outlier's threshold of the library. All the versions that lie above this line are considered significantly popular.
As shown in the figure, although the version $2.4$ of Apache IO is quite old, it is still the most popular release of this library in \mc. In the case of JUnit, it has two significantly popular versions: $4.11$ and $4.12$. On the other hand, the library XML APIs does not have any significantly popular version (\ie, the popularity of its versions remains steady across time).

\begin{figure}[htb]
	\centering
	\includegraphics[origin=c,width=0.4\textwidth]{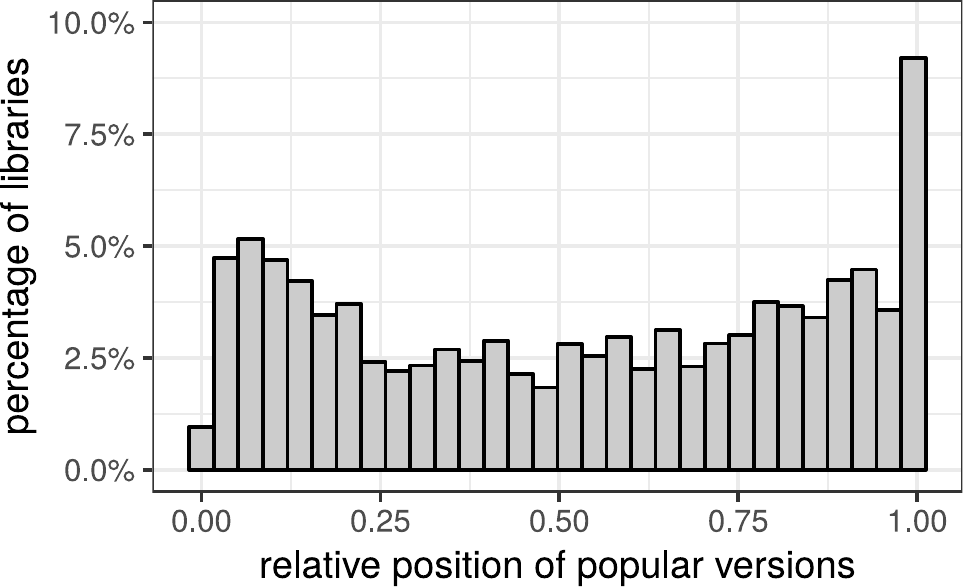}
	\caption{Histogram of the positional distribution of significantly popular library versions (\#bins = $30$).}
	\label{fig:position_most_used_version}
\end{figure}

In order to measure the positional distribution of popular versions, we focus on libraries that have at least one significantly popular version. We determine the relative position of such versions with respect to the number of versions of the library. As for the positional distribution of active version, we also normalize the relative position between $[1,0]$. The histogram in Figure~\ref{fig:position_most_used_version} shows the distribution of the positions of the most popular versions across libraries. We observe that less than $10\%$ of libraries have their latest version as the most popular. This is expected since the average lifespan of latest versions is lower than the average of non-latest versions.  Interestingly, we found that the remaining highly popular versions are almost equally distributed, between $2\%$ and $5\%$, in the remaining positions. 

This result indicates that the most popular libraries in \mc are distributed across all the different library releases. It is notable that for almost $85\%$ of libraries the most used version is not the latest. Thus, older versions are still being heavily used by other libraries, with the exception of the first version which is rarely the most popular.

\begin{mdframed}[style=mpdframe]
\textbf{Findings from RQ3:} 
$17\%$ of the libraries have more than one significantly popular version distributed across different releases, each of which creates a niche fitting a group of users. This indicates that library developers successfully address the needs of diverse populations of users.\looseness=-1
\end{mdframed}

\subsection{\RQfour}

We have seen so far that many libraries in \mc have multiple active versions, of which more than one can be significantly more popular than the others. Now, we investigate whether the activity status of versions has a direct effect on the popularity of their corresponding library. For this, we calculate the percentages of active and passive versions of each library and compare them with respect to the overall popularity of the library.\looseness=-1 

\begin{figure}[htb]
	\centering
	\includegraphics[origin=c,width=0.37\textwidth]{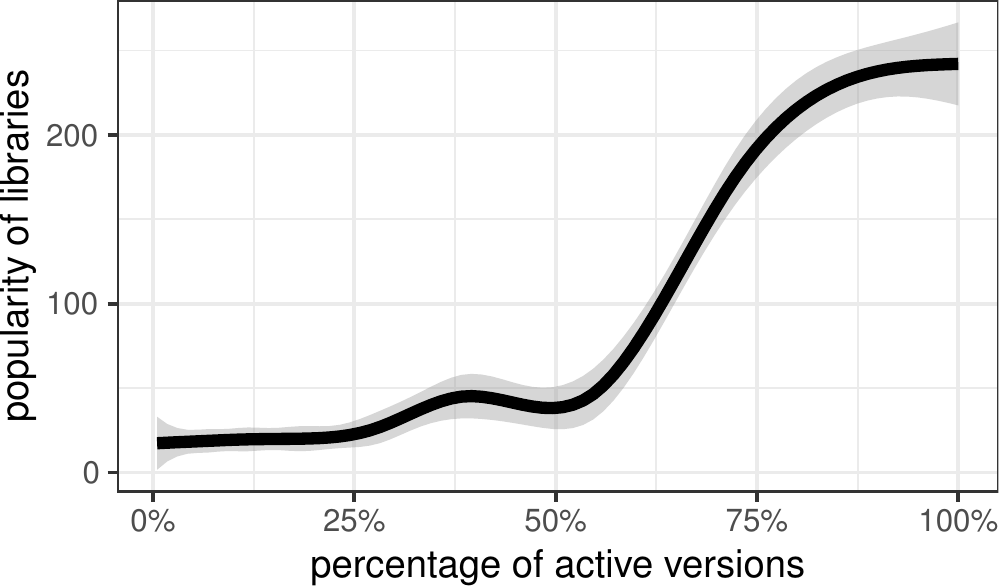}
	\caption{Fitting curve (GAM model) of the percentage of active versions \wrt the popularity of libraries ($pop_\mathcal{L}(l)$ metric). The shaded area around the fitting curve represents the $95\%$ confidence interval.
	}
	\label{fig:popularity_vs_percentage_of_active_versions}
\end{figure}

Figure~\ref{fig:popularity_vs_percentage_of_active_versions} shows the smoothing function corresponding to the relation between the popularity of libraries and their percentages of active versions. There is a significant positive correlation between both variables (Spearman's rank correlation test: $\rho = 0.87$, p-value $< 0.01$). In particular, we observe that libraries that have more than $50\%$ of active versions are more likely to be very popular, as popular libraries with many versions attract more clients for their versions.\looseness=-1 


Table~\ref{tab:top_popular_libraries} shows the seven most popular libraries ranked in decreasing order of popularity, as well as their percentage of active and significantly popular versions. As we can see, in all the cases a significant proportion of their versions are active. This indicates that many versions of these libraries continue being actively used, contributing to the popularity of the library and adding dependency diversity among all the clients. In three cases out of seven, there are more than two versions that are significantly more popular than the others. Finally, we also notice that these popular libraries serve general purposes, which allow them to fit well for various types of usages.\looseness=-1

\begin{mdframed}[style=mpdframe]
\textbf{Findings from RQ4:} Popular libraries in \mc have most of their versions active and serve general purposes. Moreover, the popularity of a library can be estimated by the number of its active versions. The more active versions a library has, the more likely it is to be popular, and vice-versa.\looseness=-1
\end{mdframed}

\begin{table}[htb]
\centering
\caption{The top-$7$ most popular libraries in our study subjects and their number of active and significantly popular versions}
\begin{tabular}{llcc} 
\toprule
\textbf{Library} & \textbf{Domain} & \textbf{\#Active (\%)} & \textbf{\#Popular (\%)} \\ 
\hline
 \emph{google.code.findbugs:jsr305} & Utility & $10$ ($90\%$) & $1$ ($9.01\%$)\\
 \emph{org.slf4j:slf4j-api} &  Logging & $63$ ($86.3\%$) &  $3$ ($4.1\%$) \\ 
 \emph{log4j:log4j} & Logging &  $18$ ($94.7\%$) & $1$ ($5.2\%$) \\ 
 \emph{com.google.guava:guava} &  Utility & $71$ ($79.7\%$) & $1$ ($1.2\%$)\\ 
 \emph{junit:junit} & Testing &  $27$ ($96.5\%$) &  $2$ ($7.1\%$)\\ 
 \emph{org.hamcrest:hamcrest-core} & Testing &  $5$ ($100\%$) & $1$ ($20\%$) \\ 
 \emph{commons-logging:logging} & Logging & $15$ ($88.3\%$) & $2$ ($11.8\%$)\\ 
\bottomrule
\label{tab:top_popular_libraries}
\end{tabular}
\end{table}

\subsection{\RQfive}

This research question focuses on the temporal dimension of the dataset. 
We analyze whether the diversity of popular and active versions that we observe today is a phenomenon that sustained in the past history of the libraries. 
We look at every single library version $v$ separately and investigate whether, during the time period when $v$ was the latest, it gained the expected attraction among its older peers.  We compare the number of usages that a version $v$ gets during its lifespan period against the number of usages that the whole library received during the timeliness period of $v$. For this comparison, we rely on the timeliness function described in Metric~\ref{met:timeliness}. This metric can be considered as an internal popularity metric that  assesses the popularity of a version among its peers. \looseness=-1

\begin{figure}[htb]
    \centering
    \includegraphics[origin=c,width=0.44\textwidth]{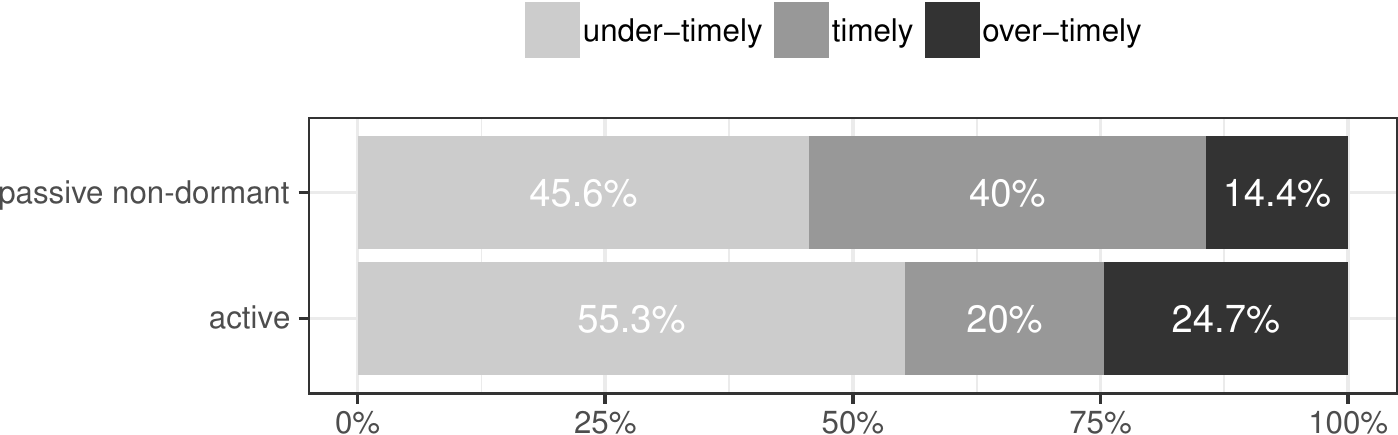}
    \caption{Proportions of timeliness classes for passive and active versions.}
    \label{fig:versions_timeliness_fill}
\end{figure}

Overall, for all our study subjects, $70.6\%$ of library versions are under-timely (including dormant versions), while $19.8\%$ are timely, and the remaining $9.6\%$ are over-timely. 
Figure~\ref{fig:versions_timeliness_fill} shows the distribution of the three timeliness classes for active and passive versions. We observe that roughly $45\%$ of passive library versions were under-timely. These are versions that did not attract users for their library throughout their timeliness period. Meanwhile, almost $55\%$ are timely. These are library versions that were not only active at some point, but also widely used. This gives substantial evidence that the diversity that we observe today has existed in the past in \mc. On the other hand, we observe that $55.3\%$ of active versions are under-timely. These are versions that are not widely popular among their peers, yet active. The average lifespan of these versions is ${\sim}777$ days, which suggests that although they are under-timely, they are likely to remain active for a long period of time; whereas, the remaining active versions are evenly distributed among timely and over-timely. 

\begin{figure}[htb]
	\centering
	\includegraphics[origin=c,width=0.35\textwidth]{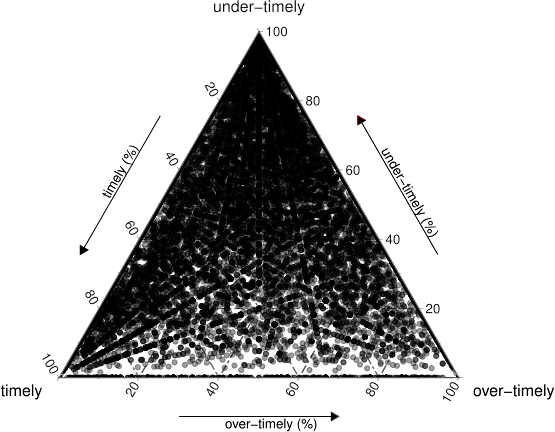}
	\caption{Distribution of libraries \wrt their percentages of over-timely, timely and under-timely versions. The dispersion of points inside the triangle indicates that the proportions of classes are well distributed across the libraries.}
	\label{fig:plot3d}
\end{figure}

In order to analyze the distribution of the timeliness classes at the library level, we calculate the proportions of under-timely, timely and over-timely versions in each library. Figure~\ref{fig:plot3d} shows a ternary diagram~\cite{Hamilton2018:Ternary_Diagrams_Using_ggplot2} representing the distribution of the three timeliness classes across the study subjects. In the figure, each point represents a library. In general, we observe a high dispersion in the space of libraries, meaning that there are representative cases for almost all of the different proportions of classes. The paired correlation tests between the proportions of each of the classes and the popularity of their corresponding library reveal that none of the correlations are statistically significant (p-value > $0.05$ according to the Spearman's test). Therefore, the proportions of the timeliness of the versions of a library are not directly related with the popularity of the library.\looseness=-1

\begin{mdframed}[style=mpdframe]
\textbf{Findings from RQ5:} 
The diversity in the usage and popularity of versions has  consistently sustained during the history of \mc. We observe that ${\sim}10\%$ of all the library versions  attract new users during their timeliness period and remain active even after the next version has been released.  Meanwhile, there is no correlation between the popularity of a library and the timeliness of its versions. 
\end{mdframed}
\section{Discussion}\label{sec:discussion}
In this section, we discuss the implications of our findings about the emergence of software diversity in \mc, as well as some threats to the validity of our results.\looseness=-1


\subsection{Supporting the Emergence of Software Diversity}

This study focuses on the diversity of usages of libraries and versions in \mc.
We have observed empirically how the immutability of versions, which is a characteristic enforced by design in \mc, supports the natural emergence of software diversity~\cite{Baudry2015:The_Multiple_Facets_of_Software_Diversity}. 
This diversity takes multiple forms and has various effects:
\begin{itemize}
    \item all active libraries have strictly more than one active version, and the 42.7\% of them have more than two active versions;
    \item 17\% of the libraries have two or more versions that are significantly more popular than the others, which indicates a very rich diversity in usages of the latest library releases and may imply that the latest library versions deployed on \mc use different versions of a similar library;
    \item the  most popular  libraries are also the ones that have the largest proportion of active versions; 
    \item the existence of multiple used versions that overlap in time is a common phenomenon in the history of all libraries.\looseness=-1
\end{itemize}
   
We interpret these multiple forms of diversity in usage and popularity of libraries as follows: a repository that offers the opportunity for users to choose their dependencies, naturally supports the emergence of diversity among these dependencies. In other words, this massive emergent diversity is not only due to users who forget to update their dependencies. Many users decide very explicitly to depend on one or the other version of a library because it perfectly fits their needs. Consequently, this kind of diversity emerges in a fully decentralized and unsupervised manner.

Our study also highlights some important challenges for a repository that supports diversification. First, there is a cost for the maintainers of \mc. We have observed that, although most libraries are actively used ($95.49\%$), only 
$14.73\%$ of the Maven artifacts are used. We have also noticed that some companies use \mc to store artifacts that nobody else uses ($45.1\%$ of versions are dormant). Consequently, keeping all versions induces an overhead in hardware and software resources. Second, there is a cost for the developers of popular libraries who need to maintain several versions of their library to serve different clients. Third, there is a risk that users decide to keep a dependency towards a vulnerable or flawed version.

The trade-off between healthy levels of diversity in a system (here, the \mc ecosystem) and the challenges of redundancy and noise is necessary and very natural. Biological studies insist on the importance of keeping less fit or even unexpressed genes as genetic material that is necessary in order to adapt to unpredictable environmental changes \cite{frankham2005:genetics, Sawant2018}. Our study reveals that the immutability of Maven artifacts provides the material for libraries to eventually fit the needs of various users, which eventually results in the emergence of diverse popular and timely versions. In the same way that biological systems do, library maintainers can accommodate the overhead of manual updates and conflict management in order to contribute to the sustainability of the massively large pool of software diversity that exists in \mc.




\subsection{Threats to Validity}\label{sec:threats}
We report about internal, external, and reliability threats to the validity of our study. 

\paragraph{Internal validity}
The internal threat relates to the metrics employed, especially those to compare the popularity of libraries and versions. In this work, we characterize popularity in terms of number of usages and quantify it based on well-known graph-based metrics~\cite{Inoue2005:Ranking_Significance_of_Software_Components}. Thus, we assume that a widely reused library is a popular one, and we consider only the relationships described in \mc, which do not take into account usages from private projects. The jOOQ library is one example among others. Because it is dual-license, many OSS libraries avoid to depend on it, but other closed-source software are still using it and there is no way to quantify their number. However, as suggested in previous studies, software popularity can be measured in a variety of ways, depending on different factors such as social or technical aspects~\cite{Zerouali2019:On_the_Diversity_of_Software_Package_Popularity_Metrics}. Another concern relates to the fact that conventions on semantic versioning are not really taken well into account by library maintainers~\cite{Raemaekers2018:Semantic_Versioning_and_Impact_of_Breaking_Changes_in_the_Maven_Repository}. Still, we believe that at the scale of the dataset employed in this study, our metrics are a fair approximation of the state of practice in \mc.\looseness=-1

\paragraph{External validity} Our results might not generalize to other software repositories beyond the \mc ecosystem (e.g., npm, RubyGems or CRAN).
It should also be noticed that \mc does not perform any real vetting of the people that deploy artifacts or on the quality of such artifacts. Thus, the integrity and origin of most of our study subjects therein is not known or verifiable. Moreover, this work takes into account version ordering as well as temporal ordering relationships, which we believe are sufficient to give a plausible representation of the way that libraries are updated as well as their evolution trends.\looseness=-1

\paragraph{Reliability validity} Our results are reproducible, the dataset used in this study is publicly available online\footnote{\url{https://doi.org/10.5281/zenodo.1489120}}. Moreover, we provide all necessary code\footnote{\url{https://github.com/castor-software/oss-graph-metrics/tree/master/maven-central-diversity}} to replicate our analysis, including Cypher queries and R notebooks.\looseness=-1


\section{Related work}\label{sec:related}
 This paper is related to a long line of previous works about mining software repositories and analysis of dependency management systems. In this section we discuss the related work along the following aspects.

\paragraph{Structure and updating behavior}
Over the past years, several research papers have highlighted the benefits of leveraging graph-based representations and ecologycal principles to analyze the architecture of large-scale software systems~\cite{Ramler2019:Benefits_and_Drawbacks_Graph_Databases,Mancinelli2016:Managing_the_Complexity_of_Large_Free_and_OSS_Packages, Mens2014:Ecological_Studies_of_Open_Source_Software_Ecosystems,Mens2015:The_Ecology_of_Software_Ecosystems}. Raemaekers et al.~\cite{Raemaekers2014:Semantic_Versioning_versus_Breaking_Changes} investigated the adherence to semantic versioning principles in \mc as well as the update trends of popular libraries. They found that the presence of breaking changes has little influence on the actual delay between the availability of a library and the use of the newer version. Kula et al.~\cite{Kula2015:Trusting_a_Library} study the latency in trusting the latest release of a library and propose four types of dependency adoptions according to the dependency declaration time. De Castilho~et al.~\cite{Castilho2014:A_Broad-coverage_Collection_of_Portable_NLP_Components_for_Building_Shareable_Analysis_Pipelines} use the \mc repository for automatically selecting and acquiring tools and resources to build efficient NLP processing pipelines. Their analysis relied partially on Maven build files to collect library dependencies in industrial systems. However, as far as we know, none of the existing works have studied the repercussion of the artifacts' immutability at the scale of the entire \mc repository.\looseness=-1

\paragraph{Analysis of evolution trends} 
The evolution of software repositories is a popular and widely-researched topic in the area of empirical software engineering. Recently, Decan et al.~\cite{Decan2018:An_Empirical_Comparison_of_Dependency_Network_Evolution_in_Seven_Software_Packaging_Ecosystems} perform a comparison of the similarities and differences between seven large dependency management systems based on the packages gathered and archived in the \emph{libraries.io} dataset. They observe that dependency networks tend to grow over time and that a small number of libraries have a high impact on the transitive dependencies of the network. Kikas et al.~\cite{Kikas2017:Structure_and_Evolution_of_Package_Dependency_Networks} study the fragility of dependency networks of JavaScript, Ruby, and Rust and report on the overall evolutionary trends and differences of such ecosystems. Abdalkareem et al.~\cite{Abdalkareem2017:Why_Do_Developers_Use_Trivial_Packages} investigate about the reasons that motivate developers to use trivial packages on the npm ecosystem. Raemaekers et al.~\cite{Raemaekers2013:The_Maven_Repository_Dataset_of_Metrics_Changes_and_Dependencies} construct a Maven dataset to track the changes on individual methods, classes, and packages of multiple library versions. Our work expands the existing knowledge in the area by showing how software repositories can contribute to prevent dependency monoculture by making available a more diverse set of library versions for software reuse.


\paragraph{Security and vulnerability risks}
Researchers have investigated and compared dependency issues across many packaging ecosystems. Suwa et al.~\cite{Suwa2017:An_Analysis_of_Library_Rollbacks} investigate the occurrence of rollbacks during the update of libraries in Java projects. Their results confirm previous studies that show that library migrations have no clear patterns and in many cases, the latest available version of a library is not always the most used~\cite{Teyton2014:A_Study_of_Library_Migrations_in_Java, Teyton2012:Mining_Library_Migration_Graphs}. Mitropoulos et al.~\cite{Mitropoulos2014:The_Bug_Catalog_of_the_Maven_Ecosystem} present a dataset composed of bugs reports for a total of $17,505$ Maven projects. They use FindBugs to detect numerous types of bugs and also to store specific metadata together with the FindBugs results. Zapata et al.~\cite{Zapata2018:Towards_Smoother_Library_Migrations} compare how library maintainers react to vulnerable dependencies based  on whether or not they use the affected functionality in their client projects. Our work considers security and vulnerability risks in software repositories from a novel perspective, i.e., by taking into account the benefits and drawbacks that come with the emergence of software diversity.\looseness=-1

\section{Conclusion}\label{sec:conclusion}

In this paper, we performed an empirical study on the diversity of libraries and versions in the \mc repository. We studied the activity, popularity and timeliness of $1,487,956$ artifacts that represent all the versions of $73,653$ libraries. We defined various graph-based metrics based on the dependencies among Maven artifacts that are captured in the Maven Dependency Graph~\cite{Benelallam2019}. We found that ${\sim}40\%$ of libraries have two or more versions that are actively used, while almost $4\%$ never had any user in \mc. We also found that more than $90\%$ of the most popular versions are not the latest releases, and that both active and significantly popular versions are distributed across the history of library versions. In summary, we presented quantitative empirical evidence about how the immutability of artifacts in \mc supports the emergence of natural software diversity, which is fundamental to prevent dependency monoculture during software reuse.
Our next step is to investigate how we can amplify this natural emergence of software diversity through dependency transformations at the source code level.\looseness=-1


\section*{Acknowledgments}\label{sec:ak}
This work has been partially supported by the EU Project STAMP ICT-16-10 No.731529, by the Wallenberg Autonomous Systems and Software Program (WASP), by the TrustFull project financed by the Swedish Foundation for Strategic Research and by the OSS-Orange-Inria project.\looseness=-1


\bibliographystyle{ieeetr}
\end{document}